\documentclass[a4paper,11pt]{article}

\usepackage{hyperref}
\usepackage{braket}
\usepackage{amssymb}
\usepackage{amsmath}
\usepackage{amsthm}
\usepackage{amscd}
\usepackage{mathptmx}
\usepackage[all,arc,knot,matrix]{xy}
\usepackage{tikz}
\usetikzlibrary{circuits.ee.IEC}
\usepackage{circuitikz}
\usepackage{fullpage}

\newcommand{\qw}[1][-1]{\ar @{-} [0,#1]}
\newcommand{\qwd}[1][-1]{\ar @{--} [0,#1]}

\newcommand{\gate}[1]{*+<.6em>{#1} \POS ="i","i"+UR;"i"+UL **\dir{-};"i"+DL **\dir{-};"i"+DR **\dir{-};"i"+UR **\dir{-},"i" \qw}
\newcommand{\gated}[1]{*+<.6em>{#1} \POS ="i","i"+UR;"i"+UL **\dir{-};"i"+DL **\dir{-};"i"+DR **\dir{-};"i"+UR **\dir{-},"i" \qwd}

\newcommand{\measure}[1]{*+[F-:<.9em>]{#1} \qw}

\newcommand{\measureD}[1]{*{\xy*+=<0em,.1em>{#1}="e";"e"+UR+<0em,.25em>;"e"+UL+<-.5em,.25em> **\dir{-};"e"+DL+<-.5em,-.25em> **\dir{-};"e"+DR+<0em,-.25em> **\dir{-};{"e"+UR+<0em,.25em>\ellipse^{}};"e"+C:,+(0,1)*{} \endxy} \qw}
\newcommand{\measuredD}[1]{*{\xy*+=<0em,.1em>{#1}="e";"e"+UR+<0em,.25em>;"e"+UL+<-.5em,.25em> **\dir{-};"e"+DL+<-.5em,-.25em> **\dir{-};"e"+DR+<0em,-.25em> **\dir{-};{"e"+UR+<0em,.25em>\ellipse^{}};"e"+C:,+(0,1)*{} \endxy} \qwd}
\newcommand{\multimeasure}[2]{*+<1em,.9em>{\hphantom{#2}} \qw \POS[0,0].[#1,0];p !C *{#2},p \drop\frm<.9em>{-}}
\newcommand{\multimeasureD}[2]{*+<1em,.9em>{\hphantom{#2}} \POS [0,0]="i",[0,0].[#1,0]="e",!C *{#2},"e"+UR-<.8em,0em>;"e"+UL **\dir{-};"e"+DL **\dir{-};"e"+DR+<-.8em,0em> **\dir{-};{"e"+DR+<0em,.8em>\ellipse^{}};"e"+UR+<0em,-.8em> **\dir{-};{"e"+UR-<.8em,0em>\ellipse^{}},"i" \qw}

\newcommand{\multimeasureDl}[2]{*+<1em,.9em>{\hphantom{#2}} \POS [0,0]="i",[0,0].[#1,0]="e",!C *{#2},
"e"+UL+<.8em,0em>;
"e"+UR **\dir{-};
"e"+DR **\dir{-};
"e"+DL+<.8em,0em> **\dir{-};
{"e"+DL+<0em,.8em>\ellipse_{}};
"e"+UL+<0em,-.8em> **\dir{-};
{"e"+UL+<.8em,0em>\ellipse_{}},}

\newcommand{\multigate}[2]{*+<1em,.9em>{\hphantom{#2}} \POS [0,0]="i",[0,0].[#1,0]="e",!C *{#2},"e"+UR;"e"+UL **\dir{-};"e"+DL **\dir{-};"e"+DR **\dir{-};"e"+UR **\dir{-},"i" \qw}
\newcommand{\ghost}[1]{*+<1em,.9em>{\hphantom{#1}} \qw}

\newcommand{\gategroup}[6]{\POS"#1,#2"."#3,#2"."#1,#4"."#3,#4"!C*+<#5>\frm{#6}}

\newcommand{\Qcircuit}{\xymatrix @*=<0em>}

\newcommand{\pureghost}[1]{*+<1em,.9em>{\hphantom{#1}}}
\usepackage[small,nohug,heads=vee]{diagrams}
\diagramstyle[labelstyle=\scriptstyle]

\theoremstyle{definition}

\theoremstyle{remark}

\newcommand{\abs}[1]{\lvert#1\rvert}

\newcommand\Ground{%
\mathbin{\text{\begin{tikzpicture}[circuit ee IEC,yscale=0.6,xscale=0.5]
\draw (-2ex,0) to (0,0) node[ground,xshift=.65ex] {};
\end{tikzpicture}}}%
}

\begin{document}

\title{S-matrix Interpretation in Categorical Quantum Mechanics}

\author{Xiao-Kan Guo}

\date{}

\maketitle

\begin{abstract}
We study the $S$-matrix interpretation of quantum theory in light of Categotical Quantum Mechanics. The $S$-matrix interpretation of quantum theory is shown to be a functorial semantics relating the algebras of quantum theory to the effective $S$-matrix formalism. Consequently, issues such as state reduction and entanglement generation can be depicted in a simple manner. Moreover, this categorical $S$-matrix interpretation does not have the alleged thermodynamic cost.
\end{abstract}

\section{Introduction}
In the last century, quantum theory has received numerous reformulations, reconstructions and interpretations. Although these studies in quantum foundations do not change the theoretical predictions of quantum theory, they shape what we can learn about Nature from the quantum phenomena we observe.  In recent years there is a revival of discussions on the quantum foundational issues inspired by the prosperity of quantum information and computation. The prospects of quantum foundations have evolved from the consistency of quantum theory to applications such as quantum information theory and even to the elusive problem of quantum gravity \cite{HS10}. In particular, interpretation is the crux of quantum foundations, for {\it``the interpretation is the link between the [mathematical] formalism and the [physical] phenomena"} \cite{Eng13}.

In this paper, we will revisit the interpretational issues of quantum mechanics using its categorical sematics.
Apart from the popular or well-known interpretations of quantum theory (cf. \cite{Cab15} and references therein), there are still many unpopular ones that deserve more attentions, one of which is the $S$-matrix interpretation proposed by H. P. Stapp \cite{Sta71}. Usually the paper \cite{Sta71} is referred to as giving a reinterpretation of Bell's theorem, or indicating the possibility of mind-matter relations in quantum mechanics, instead of the $S$-matrix interpretation itself. This is perhaps due to the later unpopularity of the $S$-matrix program. Nevertheless, the $S$-matrix interpretation is {\it pragmatic}, by which in the morden language it could mean {\it operational}, and it is in fact {\it categorical} as we shall show in this paper.

The categorical semantics of quantum mechanics  also have diverse meanings. One version is the Categorical Quantum Mechanics (CQM) \cite{AC04}. CQM is  a reformulation of quantum theory with successful applications in various topics \cite{CK}. It is interesting to notice that without the underlying categories, the diagrammatic representation of CQM is very similar to the bubble diagrams for $S$-matrices. And it  even becomes a trivial observation for everyone that the diagrams in CQM  are similar to  quantum circuits. Based on these similarities, one might ask whether there is an unifying  principle underlying these theories. An existing example is the relation between the unitary $S$-matrices of an integrable model and the universal quantum gates for a quantum computor, which rests upon the exact fractorizations in either cases \cite{Z13}. But, in general, a model of quantum processes may not be exactly soluble, in which case the effective $S$-matrix formalism nonetheless provides unitary cluster-decomposable $S$-matrices that keep those similarities. Then the $S$-matrix interpretation of  quantum theory is a good starting point of this line of investigations. For these reasons, we  reformulte the  $S$-matrix interpretation in the settings of CQM in this paper.

In the next section, we first recollect the key ingredients of the $S$-matrix interpretation of quantum theory and also the basic language of CQM as  preliminaries. In Sec.\ref{333}, we present a functorial semantics of the $S$-matrix interpretation of quantum theory and apply it to obtain simple understandings of  the issues like state reduction and entanglement. In Sec.\ref{4444}, we clarify that in this categorical framework the so-called thermodynamic cost of interpretations of quantum theory is a wrong construction, which is because here the $S$-matrices are in the torsion class of an abelian category with zero entropy production. Finally, in Sec. \ref{5555} some outlooks are given.
\section{Preliminaries}
\subsection{The  S-matrix interpretation of quantum theory} In the Copenhagen interpretation of quantum theory, one must divide the physical world into two parts: the oberved system and the observing system. The observed quantum system is represented by probability functions which depends on the microscopic interactions in the observed system, whereas the observing system is described by classical concepts such as the responses of classical devices. This separation is sketched in Tab.\ref{tab1}. However, the probability functions as functions of the degrees of freedom of the observed quantum system give probabilities of responses appearing in the classical measureing devices. Nowadays one can easily understand this quantum-to-classical connection via decoherence,\footnote{In the framework of decoherence, one also considers the quantum part $\mathcal{A}$ of the measurement apparatus to which the information is transferred from the the observed quantum system $\mathcal{S}$. Then the decoherence happens in $\mathcal{A}$ from which one reads the classical pointer-basis states. Here we simply consider the quantum-classical cut, which is safe since the information transfer between $\mathcal{S}$ and $\mathcal{A}$ is still achieved through unitary interactions.} 
but in \cite{Sta71} Stapp gives a different observation:

\begin{table}[t]
\caption{The division of the world into observed and observing systems.}
\label{tab1}
\centering
\begin{tabular}{lll}
\hline
system& d.o.f.& description\\
\hline
observed system& quantum & probability \\
observing system& classical & response \\
\hline
\end{tabular}
\end{table}
If one takes into account the full interaction (including particle creations and annihilations) between preparing and measuring devices, the number of degrees of freedom will be infinite. But the classical devices only have finite degrees of freedom. In order to extract finite data from the full interaction, one should use  $S$-matrix theory to study the (long-range) correlations between the preparing and measuring devices.

As an example, consider the simple case where the prepared state $\ket{\psi_{\text{in},A}}$ is measured in $\ket{\psi_{\text{out},B}}$, then
\begin{align*}
\braket{\psi_{\text{in},B}|S|\psi_{\text{in},A}}=&\braket{\psi_{\text{out},B}|\psi_{\text{in},A}}=\braket{\psi_{B}(t)|U^\dagger(t,-\infty) U(t,-\infty)|\psi_{A}(t)}=\braket{\psi_{B}(t)|\psi_{A}(t)}
\end{align*}
which means that the correlation between $A$ and $B$ in the Schr\"odinger picture can be described by $S$-matrices, or diagrammatically 
\[
\braket{\psi_{B}(t)|\psi_{A}(t)}=
\xymatrix{
\ar@{-}[r] &*+[F]{S}&\ar@{-}[l]
}.
\]
Notice that although we have used the asymptotic evolution operators $U(t,-\infty)$, the asymptoticity is not a necessary condition when we apply  $S$-matrices to non-relativisitic quantum theory because a scattering theory with finite (small) distance is mathematically achievable (see, e.g. \cite{LLD14}).

In general, the  $S$-matrix interpretation of quantum theory is based on the following (slightly variant version of) pragmatic descriptions:
\begin{enumerate}
\item {\it Extrinsic Descriptions of Physical Objects}. One can only describe the observed system by its effects upon the observing system.  Such a description is necessarily probabilistic, since the observing system only has access to the finitary  $S$-matrix, which is not sufficient to deterministically describe the full interaction.
\item {\it Transition Probabilities}. If $\rho_A$ is the phase-space probability density in the preparing device  and $\rho_B$ in the measuring device, then the transition probability
\[P(A,B)=\int dxdp\rho_A(x,p)\rho_B(x,p)=\abs{\braket{\psi_{B}|\psi_{A}}}^2\]
is the probability of a succesful observation in $B$ if it is prepared in $A$. The tansition probability is an intrinsic quantity of the observed states, but can be obtained from the extrinsic probability functions in the devices.
\item {\it Dynamics}. Now that we have used the extrinsic descriptions, the suitable danymical laws for  the observed  system should be the CP maps.
\end{enumerate}
These conceptual descriptions are rather vague. An explicit description can be implemented in the veteran $S$-matrix theory (cf. Appendix \ref{AAA}).

We say these descriptions are pragmatic because they conform to the pragmatist view of quantum theory: a quantum state neither represents the ontological physical reality which could include the full complicated classical-quantum interactions, nor the epistemological knowledge of a classical observer, but {\it ``offers guidance on the legitimacy and limitations of descriptive claims about a physical situation"} \cite{Hea12}, where the descriptive claims are made by some (not necessarily real) agent who can only form expectations of future claims by legitimate conditions set by the previous quantum states. Phrased in the $S$-matrix interpretation, a quantum state is used to provide incoming and outgoing ``particles" of a scattering process, and the $S$-matrices are used as an agent to calculate the probabilties of outcomes in a ``scattering" process. These ``particles" and ``scatterings" are theoretical  constructs to effectively describe the real processes whose full dynamics are very likely beyond the scope of quantum theory. There are possible situations where no real ``scattering of particles" happens but quantum theory still applies.
Meanwhile, the $S$-matrices are objective in that they give the correct statistical correlations between quantum states, thereby making the $S$-matrix interpretation both $\psi$-epistemic and $\psi$-ontic----a case that evades the dichotomic classification of  interpretations of quantum theory \cite{Cab15}.

\subsection{Basic categorical quantum mechanics} Let us  recall some basic categorical semantics of quantum mechanics \cite{AC04} and their diagrammatical representations \cite{CK}. In this paper we use, for the ease of typing\footnote{We make an apology here for the ugliness of the following diagrams.}, diagrams more akin to quantum circuits than the string diagrams used in \cite{CK}. As is used in \cite{GS17}, such circuit-like diagrams have advantages of relating CQM to operational probabilistic theories.

The basic category ${\bf C}$ consists of as objects the {\it quantum states} (or quantum systems) and as morphisms the {\it unitary evolutions} thereof.  Let us denote by a wire/line the state $A$ and by a box the unitary transformation $U$, 
\[\psi_{A}=\xymatrix{
\ar@{-}[r] &
},\quad U:
\xymatrix{
\ar@{-}[r] &*+[F]{U}&\ar@{-}[l]
}.\]
Then clearly the identity morphism $U_{\text{id}}={\bf1}$ is an effect without any box and the  unit object is the empty diagram.

Define in ${\bf C}$  the sequential composition $\circ$ and the parallel composition $\otimes$ (as in a sequential effect algebra)
\begin{align*}
U_1\circ U_2:&\xymatrix{
\ar@{-}[r] &*+[F]{U_2}&*+[F]{U_1}\ar@{-}[l]&\ar@{-}[l]
},\\
\psi_{A}\otimes\psi_{B}=&\vcenter{
\begin{xy}
(0,0)*+{}="A";
(20,0)*+{}="B";
{\ar@{-} "A";"B"};
(0,-4)*+{}="A";
(20,-4)*+{}="B";
{\ar@{-} "A";"B"};
(10,-2)*+{\otimes}="\otimes";
\end{xy}
}=\vcenter{
\begin{xy}
(0,0)*+{}="A";
(20,0)*+{}="B";
{\ar@{-} "A";"B"};
(0,-4)*+{}="A";
(20,-4)*+{}="B";
{\ar@{-} "A";"B"};
\end{xy}
}.
\end{align*}
These compositions are associative, so that ${\bf C}$  becomes a (strict\footnote{By strict it means that the monoidal functors are all idenities. Any monoidal category is categorically equivalent to a strict monoidal category. Cf. \cite{Mac98}.}) monoidal categoery. Define in ${\bf C}$  furthermore the operation of braiding
\[\sigma_{AB}:\psi_{A}\otimes\psi_{B}\rightarrow\psi_{B}\otimes\psi_{A}\]
and require it to be symmetric $\sigma_{BA}\circ\sigma_{AB}={\bf 1}$. Diagrammatically, this is simply
\[\vcenter{
\xy
\hcross~{(0,0)}{(10,0)}{(0,5)}{(10,5)};
\endxy}
\vcenter{
\xy
\hcross~{(0,0)}{(10,0)}{(0,5)}{(10,5)};
\endxy}=\vcenter{
\begin{xy}
(0,0)*+{}="A";
(20,0)*+{}="B";
{\ar@{-} "A";"B"};
(0,-4)*+{}="A";
(20,-4)*+{}="B";
{\ar@{-} "A";"B"};
\end{xy}
}.
\]
So one has a symmetric monoidal category ({SMC}). The hermitian conjugate operation defines a {\it dagger functor} $\dagger$ in such a way that to each morphism $U:A\rightarrow B$ assigns its conjugate $U^\dagger:B\rightarrow A$. If we denote a quantum ket-state (or simply {\it state}) by the following input
\[\ket{\psi}=
\xy*+=<0em,.1em>{\psi}="e";
"e"+UL+<-.25em,.25em> ;
"e"+UR+<.25em,.25em> **\dir{-};
"e"+DR+<.25em,-.25em> **\dir{-};
"e"+DL+<-.25em,-.25em> **\dir{-};
{"e"+UL+<-.25em,.25em>\ellipse_{}};"e"+C:,+(0,1)*{};
\endxy\xy(0,0)*{}="a";(10,0)*{}="b";{\ar @{-}"a";"b"}\endxy
\]
then its dagger-conjugate bra-state (or simply {\it effect}) is
\[(\ket{\psi})^\dagger=\bra{\psi}=\xy(0,0)*{}="a";(10,0)*{}="b";{\ar @{-}"a";"b"}\endxy
\xy
*+=<0em,.1em>{\psi}="e"
;"e"+UR+<0em,.25em>;
"e"+UL+<-.5em,.25em> **\dir{-};
"e"+DL+<-.5em,-.25em> **\dir{-};
"e"+DR+<0em,-.25em> **\dir{-};
{"e"+UR+<0em,.25em>\ellipse^{}};"e"+C:,+(0,1)*{} 
\endxy.
\]
The combination of state-effect gives the inner product giving rise to probabilities,
\[
\xy*+=<0em,.1em>{\varphi}="e";
"e"+UL+<-.25em,.25em> ;
"e"+UR+<.25em,.25em> **\dir{-};
"e"+DR+<.25em,-.25em> **\dir{-};
"e"+DL+<-.25em,-.25em> **\dir{-};
{"e"+UL+<-.25em,.25em>\ellipse_{}};"e"+C:,+(0,1)*{} 
\endxy\xy(0,0)*{}="a";(10,0)*{}="b";{\ar @{-}"a";"b"}\endxy
\xy
*+=<0em,.1em>{\psi}="e"
;"e"+UR+<0em,.25em>;
"e"+UL+<-.5em,.25em> **\dir{-};
"e"+DL+<-.5em,-.25em> **\dir{-};
"e"+DR+<0em,-.25em> **\dir{-};
{"e"+UR+<0em,.25em>\ellipse^{}};"e"+C:,+(0,1)*{} 
\endxy~=~\braket{\psi|\varphi}
\]
and one can write the resolution of the identity in the obvious manner
\[
{\bf 1}=~\psi~\xy(0,0)*{}="a";(10,0)*{}="b";{\ar @{-}"a";"b"}\endxy ~\psi~=~
\sum_X~\xy(0,0)*{}="a";(10,0)*{}="b";{\ar @{-}"a";"b"}\endxy
\xy
*+=<0em,.1em>{X}="e"
;"e"+UR+<0em,.25em>;
"e"+UL+<-.5em,.25em> **\dir{-};
"e"+DL+<-.5em,-.25em> **\dir{-};
"e"+DR+<0em,-.25em> **\dir{-};
{"e"+UR+<0em,.25em>\ellipse^{}};"e"+C:,+(0,1)*{} 
\endxy~
\xy*+=<0em,.1em>{X}="e";
"e"+UL+<-.25em,.25em> ;
"e"+UR+<.25em,.25em> **\dir{-};
"e"+DR+<.25em,-.25em> **\dir{-};
"e"+DL+<-.25em,-.25em> **\dir{-};
{"e"+UL+<-.25em,.25em>\ellipse_{}};"e"+C:,+(0,1)*{} 
\endxy\xy(0,0)*{}="a";(10,0)*{}="b";{\ar @{-}"a";"b"}\endxy.
\]
Therefore ${\bf C}$  becomes a dagger symmetric monoidal category ($\dagger$-SMC). \footnote{Beside the $\dagger$-conjugate, there are more fine-grained notions like transpose and adjoints. Cf. \cite{CK}.}

The $\dagger$-SMC is defined for a unipartite quantum system. When generalized to bipartite quantum systems, it contains more strucutres such as non-separable states and subsystems. First, introduce the morphisms $\eta_A:{\bf1}\rightarrow A\otimes A$ and $\epsilon_A:A\otimes A\rightarrow{\bf1}$ such that,
\[(\epsilon_A\otimes{\bf1}_A)\circ({\bf1}_A\otimes\eta_A)={\bf1}_A,\quad\epsilon_A\circ\sigma_{AA}=\epsilon_A,\quad\sigma_{AA}\circ\eta_A=\eta_A,\]
which can be taken as the characterizations of bipartite non-separable states. Diagrammtically, if we denote $\eta$ and $\epsilon$ respectively by
\[\eta:~\vcenter{
\Qcircuit @C=1em @R=.7em {
& \multimeasureDl{1}{\text{}} &\qw\\
& \pureghost{\text{}}&\qw
}}\quad,\qquad
\epsilon:~\vcenter{
\Qcircuit @C=1em @R=.7em {
& \multimeasureD{1}{\text{}} \\
& \ghost{\text{}}
}}
\]
then they satisfy the following conditions
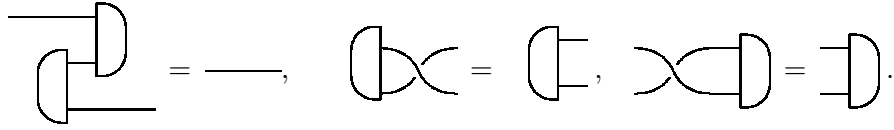
\[\vcenter{
\Qcircuit @C=1em @R=.7em {
&\qw& \multimeasureD{1}{\text{}} \\
&\multimeasureDl{1}{\text{}}& \ghost{\text{}}\\
&\pureghost{\text{}}&\qw&\qw
}}~=~\xy(0,0)*{}="a";(10,0)*{}="b";{\ar @{-}"a";"b"}\endxy,\quad
\vcenter{
\Qcircuit @C=1em @R=.7em {
& \multimeasureDl{1}{\text{}} \\
& \pureghost{\text{}}
}}
\vcenter{
\xy
\hcross~{(0,1)}{(10,1)}{(0,7)}{(10,7)};
\endxy}~=\vcenter{
\Qcircuit @C=1em @R=.7em {
& \multimeasureDl{1}{\text{}} &\qw\\
& \pureghost{\text{}}&\qw
}}~,
\quad
\vcenter{
\xy
\hcross~{(0,1)}{(10,1)}{(0,7)}{(10,7)};
\endxy}
\vcenter{
\Qcircuit @C=1em @R=.7em {
& \multimeasureD{1}{\text{}} \\
& \ghost{\text{}}
}}~=~\vcenter{
\Qcircuit @C=1em @R=.7em {
& \multimeasureD{1}{\text{}} \\
& \ghost{\text{}}
}}~.
\]
Note that one can take following combination to designate {\it trace}
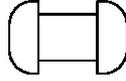
\[
\Qcircuit @C=1em @R=.7em {
& \multimeasureDl{1}{\text{}} &\qw&\multimeasureD{1}{\text{}}\\
& \pureghost{\text{}}&\qw& \ghost{\text{}}
}
\]
A trace is partial if only a subsystem is traced, e.g.
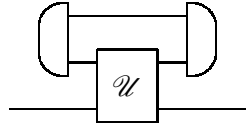
\[
\Qcircuit @C=1em @R=.7em {
& \multimeasureDl{1}{\text{}} &\qw&\multimeasureD{1}{\text{}}\\
& \pureghost{\text{}}&\multigate{1}{\mathcal{U}}& \ghost{\text{}}\\
&\qw&\ghost{\mathcal{F}}&\qw&\qw
}
\]
Obviously one has $\eta^\dagger=\epsilon$, and hence ${\bf C}$  becomes a dagger-compact closed category ($\dagger$-CCC).

Next, in order to describe more general systems interacting with classical environments, one needs to use CP maps. However, in $\dagger$-SMC (or $\dagger$-CCC) one cannot ensure the probabilities to be positive. A way around this issue is by ``doubling everything" so that the doubled wires are quantum types and the single wires are classical types (cf. \cite{CK}).
Here we assume that the single wire used above are quantum, and use dashed wires to represent a classical type $X$:
\[X~\vcenter{
\xy(0,0)*{}="a";(15,0)*{}="b";{\ar @{}"a";"b"};
(0,1.5)*{}="a";(15,1.5)*{}="b";{\ar @{--}"a";"b"}\endxy}~X.\]
 The discarding (of a subsystem) into the environment is represented by
\[
\begin{tikzpicture}[circuit ee IEC]
\draw (0,0) -- (2,0) node [ground, right] {};
\end{tikzpicture}
\]
Then a pure state does not involve any $\Ground{}$, and hence $\Ground{}$ represents an environment structure. Therefore we can represent a CP map by a mixing map
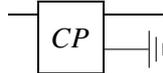
\[
\Qcircuit @C=1em @R=.7em {
& \multigate{1}{CP} & \qw &\qw\\
& \pureghost{CP}& \begin{tikzpicture}[circuit ee IEC]
\draw (2.4,0) -- (3,0) node [ground, right] {};
\end{tikzpicture}
}
\]
We call the new $\dagger$-CCC with as morphisms the CP maps a category of completely positive maps (CPM).\footnote{At this point, one can understand the necessity of $\dagger$-SMC {\it a posteriori} by comparing them with quantum channels. In particular, the composition $\circ$ corresponds to the concatenation of channels, $\otimes$ to tensor products, and $\dagger$ to conjugate of a channel.} 
The environment structure $\Ground{}$ is a terminal object of CPM, which physically means that if each system has an effect $\Ground{}$ the process is causal.

Moreover, processes involving classical types can be effectively described by ``spiders" \cite{CK}. Here again let us use circuit-like diagrams. Two basic features of classical types are copying and deleting as a comonoid:
\[\delta:A\rightarrow A\otimes A:
\xymatrix{
\ar@{--}[r] &*+[F]{\text{copy}}&\ar@2{--}[l]
},
\qquad\tau:A\rightarrow{\bf1}:
\xymatrix{
\ar@{--}[r] &*+[F]{\text{delete}}&
}
\]
Together with their $\dagger$-conjugate monoids, one can form a dagger-special commutative Frobenius algebra ($\dagger$-SCFA).\footnote{They satisfy the following conditions: $(\delta)^\dagger\circ\delta={\bf1},~(\delta)^\dagger\circ\sigma_{AA}=(\delta)^\dagger$ and the Frobenius equations. Their diagrams are obvious (Cf. \cite{CK}).} A $\dagger$-hypergraph category ($\dagger$-HC) is a category each object of which has a $\dagger$-SCFA. Furthermore, if a $\dagger$-SCFA satisfies $\tau\circ(\delta)^\dagger\circ\sigma_{AA}=\tau\circ(\delta)^\dagger$, we say it is a dagger-special symmetric Frobenius algebra ($\dagger$-SSFA). A CPM with objects being $\dagger$-SSFA is denoted by ${\bf CP}^*$.

With these we can define basic processes like 
\begin{equation}\label{pm}
\text{preparation}:
\xymatrix{
\ar@{--}[r] &*+[F]{{p}}&\ar@{-}[l]
},\quad
\text{measurement}:\xymatrix{
\ar@{-}[r] &*+[F]{{m}}&\ar@{--}[l]
},
\end{equation}
and decoherence
\begin{equation}\label{de}
\xymatrix{
\ar@{-}[r] &*+[F]{m}&*+[F]{p}\ar@{--}[l]&\ar@{-}[l]
}.
\end{equation}
\section{$S$-matrix Interpretation }\label{333}
In this section we  turn to the main result of this paper: the categorical semantics of the $S$-matrix interpretation of quantum theory. The categorical counterpart of the $S$-matrix interpretation is a functorial semantics interpreting the operationally obtained algebraic structure of quantum theory in a monoidal category.
\subsection{$S$-matrix Interpretation in CQM}
With the language of CQM in mind, the categorical semantics of the pragmatic $S$-matrix descriptions becomes immediate. To see this, let us first construct several basic processes.

In the extrinsic description of physical objects one only has access to the probabilities in the observing system.  The probability of an output $\chi$ (or response) in the measurement device with the prepared state being $\rho$ can be represented as
\[P(\rho,\chi)=\vcenter{
\Qcircuit @C=1em @R=.7em {
& \multimeasureDl{1}{\rho} &\qw&\gate{m}&\qwd&\measuredD{\chi}\\
& \pureghost{\rho}& \begin{tikzpicture}[circuit ee IEC]
\draw (1.4,0) -- (2,0) node [ground, right] {};
\end{tikzpicture}
}}
\]
But the preparing system is also classical, hence there must be a preparation preceding $\rho$. We can consider the following diagram
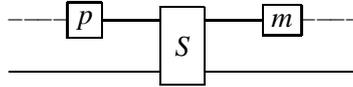
\[
\Qcircuit @C=1em @R=.7em {
&\qwd&\gated{p}&\qw& \multigate{1}{{S}}  &\qw&\gate{m}&\qwd&\qwd\\
&\qw&\qw&\qw& \ghost{{S}}&\qw&\qw&\qw&\qw
}
\]
where input state, output effect and the environment structures are hidden. We can group all the transformations into a single one
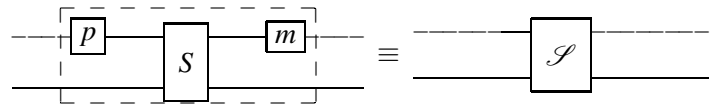
\[\vcenter{\Qcircuit @C=1em @R=.7em {
&\qwd&\gated{p}&\qw& \multigate{1}{{S}}  &\qw&\gate{m}&\qwd&\qwd\\
&\qw&\qw&\qw& \ghost{{S}}&\qw&\qw&\qw&\qw
\gategroup{1}{3}{2}{7}{.7em}{--}}}
~\equiv~\vcenter{
\Qcircuit @C=1em @R=.7em {
&\qwd&\qwd&\qwd& \multigate{1}{\mathcal{S}}  &\qwd&\qwd&\qwd&\qwd\\
&\qw&\qw&\qw& \ghost{\mathcal{S}}&\qw&\qw&\qw&\qw
}}
\]
This is a diagram representing a process with classical control. We thus see that the extrinsic descriptions via probabilties in the observing systems, in effect, are classical controls. The measuring  effects can only give  output probabilities of the classical responses since the extrinsic descriptions only constitute a subprocess as can be seen from the diagram. The total evolution is thus describable by a CP map.

Meanwhile, the classical process interacts with the  quantum system which in the above diagram is represented by the subdiagram in the dashed box. Is this quantum system the observed system? If we still include the environment structure, this quantum system would be discarded. In fact, the observed system is an effective environment to the observing system due to the incomplete knowledge  in $\mathcal{S}$, if the total system is an isolated pure process (or a lazy state with zero entropy rate). 
We therefore do not include $\Ground{}$ and consider the total state and effect
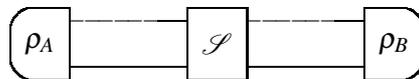
\[
\Qcircuit @C=1em @R=.7em {
& \multimeasureDl{1}{\rho_A}&\qwd&\qwd&\qwd& \multigate{1}{\mathcal{S}}  &\qwd&\qwd&\qwd& \multimeasureD{1}{\rho_B}\\
& \pureghost{\rho_A}&\qw&\qw&\qw& \ghost{\mathcal{S}}&\qw&\qw&\qw& \ghost{\rho_B}
}
\]
Then the transition probabilities can be obtained by tracing out the classical part
\begin{equation}\label{tp}
P(A,B)=\text{tr}\rho_A\rho_B=\quad
\vcenter{\Qcircuit @C=1em @R=.7em {
& \multimeasureDl{1}{\text{A}} &\qwd&\qwd&\qwd&\multimeasureD{1}{\text{B}}\\
& \pureghost{\text{A}}&\qwd&\multigate{1}{\mathcal{S}}&\qwd& \ghost{\text{B}}\\
&\xy*+=<0em,.1em>{\psi}="e";
"e"+UL+<-.25em,.25em> ;
"e"+UR+<.25em,.25em> **\dir{-};
"e"+DR+<.25em,-.25em> **\dir{-};
"e"+DL+<-.25em,-.25em> **\dir{-};
{"e"+UL+<-.25em,.25em>\ellipse_{}};"e"+C:,+(0,1)*{};
\endxy\xy(0,0)*{}="a";(10,0)*{}="b";{\ar @{-}"a";"b"}\endxy&\qw&\ghost{\mathcal{S}}&\qw&\qw&\measureD{\varphi}
}}
\end{equation}
These give us the pragmatic descriptions of quantum theory in a $\dagger$-HC (or simply CPM).

The transformation $\mathcal{S}$ here can be interpreted as an $S$-matrix whose pole sigularities are the quantum systems (solid lines). The process $\psi\rightarrow\varphi$ is then an intermediate system designated to define the $S$-matrix. Effectively, one can envision the overall process as the following $S$-matrix diagram
\[
\Qcircuit @C=1em @R=.7em {
&\xy*+=<0em,.1em>{\psi}="e";
"e"+UL+<-.25em,.25em> ;
"e"+UR+<.25em,.25em> **\dir{-};
"e"+DR+<.25em,-.25em> **\dir{-};
"e"+DL+<-.25em,-.25em> **\dir{-};
{"e"+UL+<-.25em,.25em>\ellipse_{}};"e"+C:,+(0,1)*{};
\endxy\xy(0,0)*{}="a";(10,0)*{}="b";{\ar @{-}"a";"b"}\endxy&\qw&\gate{\mathcal{S}}&\qw&\qw&\measureD{\varphi}
}
\]
which illuatrates the advantage of $S$ matrices that they can extract simple quantum processes from the complicated classical-quantum interactions. The classical observing system can then be envisioned as providing the asymptotic free sources or targets for the $S$ matrices. Hence, in the $S$-matrix description, it is not required that there must be a clear quantum-classical (Heisenberg's) cut, which is in effect replaced by hiding the classical observing system behind the $S$-matrix, but at the same time it is linked  to the observed quantum system again via the $S$-matrix. 

A clarifying consideration is that the soft modes, which can arise in computing the $S$-matrix of a scattering involving massless particles, will not jeopardize the $S$-matrix interpretation of quantum theory, since the scale of those soft particles is by definition beyond the detection scale of the detector. In effect, a measurement on the soft particles will not produce any classical responses,
\[\xymatrix{
\ar@{-}[r] &*+[F]{{m}}&
}\]
Therefore one cannot extract knowledge of the quantum process from the classical responses in this case and the extrinsic pragmatic description fails. 
This exemplifies how the full interaction in quantum field theories is coarsened to an effective $S$-matrix in quantum mechanics. In a sense the $S$-matrix interprettion hides the classical observers and forgets the higher ernergy behaviours decribed by quantum field theories.

Categorically speaking, the $S$-matrix interpretation of quantum theory  simplifies the full processes to  a pragmatic category ${\bf Set}$ of sets whose objects are sets of quantum states prepared/measured by some observing systems and morphisms are $S$-matrices relating these states. The full processes, however, form the category CPM. Then there is an ``underlying set" functor $\mathfrak{U}:\text{CPM}\rightarrow{\bf Set}$ making CPM a concrete category. In order that the full processes can be described by an algebra (of type) {\bf A}, it is necessary and sufficient that there exists a functor $\mathfrak{S}$ such that the functor $f:{\bf A}\rightarrow\text{CPM}$ is an equivalence of categories where ${\bf A}=\mathfrak{U}\hat{\mathfrak{S}}$ with $\hat{\mathfrak{S}}$ being the adjoint functor of $\mathfrak{S}$. The functor $\mathfrak{S}$ is called a {\it functorial semantics} \cite{Law63} whose codomain is the category of underlying set functors $\mathfrak{U}$ and domain is the dual of the category of algebraic theories:
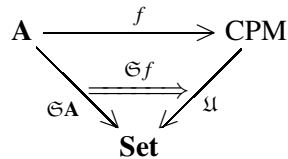
\begin{diagram}
{\bf A}& &\rTo^{f}&&\text{CPM}\\
&\rdTo_{\mathfrak{S}{\bf A}} &\rImplies^{\mathfrak{S}f}&\ldTo_{\mathfrak{U}} \\
& &\text{\bf Set}
\end{diagram}
Therefore the $S$-matrix interpretation of quantum theory plays the role of a functorial semantics from the category of algebras of quantum mechanics, say $C^*$-algebras, to the category of the forgetful concrete functors $\mathfrak{U}:\text{CPM}\rightarrow{\bf Set}$. 

Notice that here CPM is a SMC, thereby requiring a monoidal semantics \cite{Mau17}. A monoidal semantics specifies the structural rules of the term calculus in addition to the types and functional constants of an ordinary functorial semantics. For a SMC modelled on a category {\bf Hilb} of finite dimensional Hilbert spaces $\mathcal{H}$, the structure rules can be assigned as for the category {\bf Vect} of vector spaces. Namely,
\[\text{weakening}:\Delta:\mathcal{H}\rightarrow \mathcal{H}\otimes \mathcal{H},\quad\text{contraction}:\pi:\mathcal{H}\rightarrow1\]
and the exchange rule is assigned by the symmetric property of SMC. Therefore the more suitable structure could be a (cocommutative) $C^*$-Hopf algebra.

\subsection{State change}
As a consequence of the functorial semantics, there does not exist the problem of wave function collapse. There is only the change of wave function due to the change in the observed system induced by the change in the preparing settings or specifications made by the observing system. For if the measurement changes the specifications of the preparation of the observed system and results in a new state, the other structures in CPM should be changed accordingly to preserve the algebraic structure. In other words, it is a ``bookkeeping device for updating the description" \cite{Eng13}.

The selective state change or state reduction is in fact independent of the amplification in the measurement process \cite{Oza97}, which means  the state reduction can be completely characterized in the effective $S$-matrix picture. The nonselective state change, however, relies on the classical observing system. In the preceeding subsection, we have used the nonselective state change to calculate the transition probabilities \eqref{tp} in the extrinsic descriptions of quantum theory. Suppose the original state is $\rho$, and after the change the state becomes $\rho'$ which equals $\sum_aP(a)\rho_a$ for outcomes $\{a\}$ in case the change is selective.
Then the state transformation $T$ corresponding to the nonselective state change is equivalent to the sum of the the state transformation $T_a$ of the state reduction $\rho\rightarrow\rho_a$,
\begin{equation}\label{ttt}
T=\sum_aT_a.
\end{equation}
If we represent the state transformations by $S$-matrices, \eqref{ttt} becomes  the sum of reduced $S$-matrices \cite{Sad15}. We thus see that either the seletive or the nonselective state change works in the $S$-matrix interpretation.

\subsection{Entanglement by virtual scattering}
In CQM, an entangled state is represented by a non-$\otimes$-separable diagram
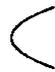
\[
\xy
(6,4)*{}="1";(6,-4)*{}="2";
**\crv{(-5,0)}
?(.20)*\dir{}+(2,-1)
?(.89)*\dir{}+(-2,-1);
\endxy 
\]
Now the problem is how two states become entangled if they are initially separated. A simple explaination is that they interact with each other, which is previously depicted by quantum spiders. Here we can simply describe it in terms of $S$-matrices:
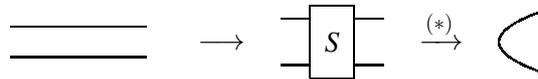
\[
\vcenter{
\begin{xy}
(0,0)*+{}="A";
(20,0)*+{}="B";
{\ar@{-} "A";"B"};
(0,-4)*+{}="A";
(20,-4)*+{}="B";
{\ar@{-} "A";"B"};
\end{xy}
}\quad\longrightarrow\quad\vcenter{
\Qcircuit @C=1em @R=.7em {
& \multigate{1}{S} & \qw \\
& \ghost{S}& \qw 
}}\quad\xrightarrow{(*)}\quad\vcenter{
\xy
(6,4)*{}="1";(6,-4)*{}="2";
**\crv{(-5,0)}
?(.20)*\dir{}+(2,-1)
?(.89)*\dir{}+(-2,-1);
\endxy }
\]
where the $(*)$ is the following
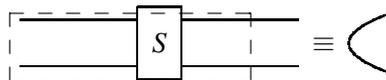
\[\vcenter{\Qcircuit @C=1em @R=.7em {
&\qw&\qw&\qw& \multigate{1}{{S}}  &\qw&\qw&\qw&\qw\\
&\qw&\qw&\qw& \ghost{{S}}&\qw&\qw&\qw&\qw
\gategroup{1}{1}{2}{7}{.7em}{--}}}
~\equiv~\vcenter{
\xy
(6,4)*{}="1";(6,-4)*{}="2";
**\crv{(-5,0)}
?(.20)*\dir{}+(2,-1)
?(.89)*\dir{}+(-2,-1);
\endxy }
\]
Here the scattering of particles needs not be a real process, since the $S$-matrix interpretation is only a prescription to understand the dynamics of quantum states within quantum theory. 

To see the quantumness of an $S$-matrix representation, we can consider the decoherence map of the  scattering system to classical theory \cite{RSA17}. If there is no entanglement in the $S$-matrix, then from \cite{RSA17} we know that the decoherence will simply discard the quantum part and result in a trivial classical state. But in general a decoherence results in a quantum state as in \eqref{de}. Resulting in a classical theory with outcome is akin to the measurement in \eqref{pm} which contradicts the decoherence defined in CQM. Therefore an $S$-matrix implies quantumness.

The simple diagrams above illustrate an important point: the quantumness appeared as the interference phases can be effectively described by a two (or more) ``particle" $S$-matrix. This allows us to go  beyond the intuition of double slits and interferometers so as to define a  measure of quantumness. For instance, the nontrivial amplitude given by the effective $S$-matrix measures entanglement and be used to derive Bell ineqaulity (see Appendix \ref{BELL}). This virtual $S$-matrix approach to entanglement is also adopted in \cite{DML09}. In a sense , this is also similar to the phase spiders in CQM \cite{CK}.

\section{Thermodynamic cost?}\label{4444}
We can now discuss some topics that have not be considered in the CQM literature with the help of the $S$-matrix interpretation. A direct issue concerning the $S$-matrix interpretation itself is its  thermodynamic cost (possibly different from other interpretations) \cite{CGGLW16}. We will see that the so-called thermodynamic cost of interpretations does not exist in our categorical framework of $S$-matrix interpretation.

The preparation-measurement pair considered above, either described by CPM or by an $S$-matrix, is in the form of an input-output process. Indeed, given a set of specifications on the preparing device, the preparation $\mathsf{P}$ is a map from the classical specifications $\mathsf{s}$ to a quantum state $\rho_{\mathsf{s}}=\mathsf{P}(\mathsf{s})$. Likewise, the measurement is a map $\mathsf{M}$ from a quantum state $\rho_{\mathsf{m}}$ to the classical responses $\mathsf{r}=\mathsf{M}(\rho_{\mathsf{m}})$. The evolution from $\rho_{\mathsf{s}}$ to $\rho_{\mathsf{m}}$ is described by either a CP map in generic cases or by a unitary $S$-matrix $S_{\mathsf{sm}}$ in our case. The quadruples $(\mathsf{P},\mathsf{M},\{\rho\},S)$ form a quantum model of an input-output process analogous to that defined in \cite{TGVG17}.

This general process structure is obviously shared by CQM, general probabilistic theories, and also computational mechanics \cite{BC15}. In computational mechanics, one is concerned with quantifying the structures of a stochastic process and optimizing the predictions. Given an input distribution of probabilities $\{p_i\}$, one can calculate its entropy $H(p_i)=-\sum_ip_1\log p_i$ as a measure of complexity of the inputs and evaluate the mutual information $I(p,p')$ with respect to the (time-)evolved future probability distribution $\{p_i'\}$ as a measure of prediction. The extrinsic description of quantum theory is an example of such a state machine. In \cite{CGGLW16} the thermodynamical cost of a finite state machine is calculated.\footnote{Note that by considering the ideal experiment of \cite{CGGLW16} as a finite state machine one cannot conclude that those interpretations are imcompatible with the assumptions. The Landauer principle gives only the lower bound of the dissipation of heat, and it is not contradictory that the heat is unbounded when the experiment repeated infinitely many times. In fact, erasing the information of previous results will make the machine less predictive, that is, one loses information about what process is going to happen, which will cause a  large amount of dissipation of heat with lower bound much greater than $kT\ln2$ (cf. \cite{GKLMSSTWL17}). The problem is that the finite state machine used in \cite{CGGLW16} is still an extrinsic description of quantum theory, for they have used the previous measurement results as the ``intrinsic" input. Hence the information is stored in the finite classical devices instead of the observed quantum system. The dissipation is still of the classical devices and has nothing to do with the assumption that the observed system is finite. A similar point of view has been put forward in \cite{PT17}.} Then is there any thermodynamical cost in the $S$-matrix interpretation?

Now in our case of quantum theory, each state $\rho$ gives the von Neumann entropy $S=-\text{tr}(\rho\log\rho)$, but the unitary quantum evolution preserves the von Neumann entropy. In other words, there is no information lost in an isolated quantum system. The situation seems more complicated if we include the interactions with the classical preparing and measuring systems. In general, a quantum operation (i.e. a trace-preserving CP map) is depolarizing and increases the entropy. But the $S$-matrix interpretation of quantum theory extracts the quantum $S$-matrix  from the quantum-classical interactions. The unitarity of $S$-matrices then ensures that the entropy is preserved. This also can be seen from the cluster decomposition of the $S$-matrix (see Appdendix \ref{BBB}). Thus, there is no thermodynamic cost in the $S$-matrix interpretation. 

Let us come back to the category {\bf Qio} of the quantum model $(\mathsf{P},\mathsf{M},\{\rho\},S)$ of input-output processes and give a categorical counterpart of the above argument. Suppose that {\bf Qio} is an abelian category. Indeed, if the evolutions are described by unitary $S$-matrices, they clearly form a unitary group. Then the assumption requires this group to be abelian , which means the possible effects of the phase factors are neglected. Then {\bf Qio} is an $Ab$-category since there exist zero objects as the empty states and direct sum decomposition as  used in the cluster decomposition. The kernel $K$ of an arrow $F$ in an abelian category is the equivalent class of monics such that $FK$ is a zero arrow, which in {\bf Qio} can be interpreted as a particular type of causal states in a $\epsilon$-transducer \cite{BC15} such that any future state will not produce any outcome. The reason why we assume the abelianness is that we can define entropy function and study the entropy of flows in it. Let $h$ be an entropy function in {\bf Qio} as defined in \cite{DB13}. Then the flows $\Phi_S$ of unitary $S$-matrices belong to the torsion class in the Pinsker torsion theory
\[\mathcal{T}_h=\{\Phi_S:{\bf P}(\Phi)={\Phi}\}\]
where the Pinsker radical is defined on an object $\rho$ of {\bf Qio} as
\[{\bf P}_h=\sum\{\sigma_i:\sigma_i\text{ is a subobject of }\rho,h(\sigma_i)=0\}.\]

The remaining torsion-free class includes those depolarizing quantum operations. Notice that the depolarizing quantum operations are  usually used to model decoherence. Then simply updating the wavefunctions does not suffice to interpret it. Categorocally, it is possible that this situation corresponds to the deviation from algebraic exactness (or from being monadic) of some 2-category. An example is the 2-category of varieties.

The next question is whether the $S$-matrix interpretation itself has thermodynamic costs. Since the underlying set functor $\mathfrak{U}$ is forgetful, there will be a change in the  entropy of information-bearing degrees of freedom. Indeed, if we consider {\bf Qio} as an information transducer that transforms the information encoded in the initial state prepared with preparing specifications to that in the final state decoded by a measurement, then there will be a transient dissipation caused by the transducer itself \cite{BMRC17}. Here in {\bf Qio} the classical preparing-measuring pairs  play the role of an information ratchet, the quantum observed system is an information-carrier, and the $S$-matrix performs the computation. But the $S$-matrix interpretation ``forgets" (or hides) the classical observing devices and keeps the quantum states and the dynamics of them. So different classical devices of the observing systems could correspond to the same quantum $S$-matrix via the $S$-matrix interpretation, which in general induces different transient dissipations. But these different experimental implementations do not affect the  quantum theory picked out by the $S$-matrix interpretation. It is thus within our expectations that the quantum theory remains intact no matter how complex the classical observing system is.
\section{Outlooks}\label{5555}
In this paper, we have made a modest contribution to CQM by reformulating the $S$-matrix interpretation in the settings of CQM. There are of course many aspects of quantum foundations that  have not been covered in this work, some of which can be approached nicely in CQM by illustrative diagrams. The $S$-matrix interpretation allows us to think about the more traditional physical pictures when drawing these diagrams.

$S$-matrices are the central concepts in quantum field theories. And there is a great progress in exporing  new relations for scattering amplitudes. It is therefore important to extend the above considerations to quantum field theories. As early steps one must develop the infinite dimensional CQM and describe spacetime causality in CQM, which are  hot topics of recent investigations. The categorical counterpart of the $S$-matrix interpretation is shown to be a functorial semantics, which in a broader sense allows the definition of existential and universal quantifications so that the functorial semantics is an interpretation of the ``first-order language" (as quantum theory) in the (quantum) ``world"  \cite{Kock71}. Then higher-order languages are natural ways to construct infinte dimensional CQM and quantum field theories, as in the non-standard analysis approach of Gogioso and Genovese \cite{GG17}.

Apart from those categorical aspects, the  physical picture that  the $S$-matrix interpretation arrives at, though unconventional, resembles the local-time scheme \cite{JAD}. The local-time scheme is also built on the asymptotic completeness of the many-body scattering theory where the so-called local times are defined or measured on each of the decomposed clusters. We can ask {\it whether} (1) composing the local times into a global time in equivalent to the composition of diagrams in CQM, and  (2) the emergence of open system dynamics from the closed scattering systems via coarse-graining is related to what we have done here to treat the pragmatic descriptions of quantum theory. CQM, as a general process theory, treats the concept of time  implicitly as relational time. The local time scheme not only supports the relational time, but also provides a calculable scheme to discuss many foundational problems. We thus hope that, through the local-time scheme, the $S$-matrix interpretation of quantum theory revisited in this paper will be able to return to its original initiative, that is, the pragmatic calculations or experimental observations.\\

{\bf Acknowledgements}. I thank Miroljub Duji\'c for helpful suggestions. This work is supported in part by the National Natural Science Foundation of China under Grant No. 61471356.

\appendix
\section{Bubble Diagrams for $S$-matrix Theory}\label{AAA}
In quantum field theories, the scattering of particles can be studied via perturbative Feynman diagrams which effectively (through the LSZ reduction formula) represent elements of the $S$-matrix. In the early days of $S$-matrix theory, however, there is another diagrammatical representation, i.e. the bubble diagrams (see for example \cite{CS69}), where the $S$-matrix itself is depicted explicitly.

Consider a scattering process represented by a Feynman diagram $D$ with vertices $\{V_i\}$ and lines $\{L_j\}$. The bubble diagram is obtained by replacing the vertices by bubbles representing the $S$-matrices and the propogators on the internal lines by $S_\alpha^{-1}$ (the inverse of the $S$-matrix $S$ restricted to the space of lines $\alpha$). More explicitly, the connected part of an $S$-matrix is represented by a (rightward-directed)  bubble with incoming and outgoing lines
\[\Qcircuit @C=1em @R=.7em {
& \measure{+} & \qw
}\]
where the plus sign inside the bubble denotes the rightward-directedness of the $S$-matrix whose conjugate  is denoted by a minus sign. Just like Feynman diagrams, one should integrate over all internal lines and sum over all ``particle" types $t_j$ to obtain the scattering ammplitude from a bubble diagram $B$
\[M^B(K;K')=\sum_{t_j}\int\frac{dp^4_j}{(2\pi)^4}2\pi\theta(p^0_j)\delta(p_j^2-\mu^2(t_j))\prod_jS_j^{-1}\prod_iS_i
\]
where $(K;K')$ are the data of external lines. A general $S$-matrix can be decomposed into connected bubble diagrams via the cluster decomposition
\[S(K;K')=\sum_BM^B(K;K')\]
where the sum is over those $B$'s with plus signs and with no internal lines. Diagrammatically, this can be sketched as
\[\xymatrix{
\ar@{.}[r] &*+[F]{+}&\ar@{.}[l]&=&\sum\ar@{.}[r] &*+[o][F-]{+}&\ar@{.}[l]
}
\]
where the sum is over all different diagrams including  bubbles and the dotted line represents all possible external lines. Then the unitarity of the $S$-matrix is simply
\[\xymatrix{
\ar@{.}[r] &*+[F]{+}&*+[F]{-}\ar@{.}[l]&\ar@{.}[l]}=\xymatrix{
\ar@{.}[r] &}\]
where the summation of all possible middle lines is taken. The discontinuity around a physical-region singularity is represented by
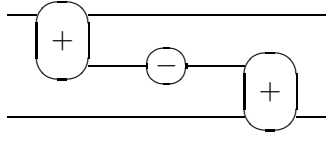
\[
\Qcircuit @C=1em @R=.7em {
& \multimeasure{1}{+} & \qw& \qw& \qw & \qw&\qw\\
& \pureghost{+}& \qw &\measure{-}&\qw& \multimeasure{1}{+}\\
& \qw& \qw &\qw&\qw& \ghost{+}&\qw
}
\]

The physical-region singularities can be interpreted as on-shell particles and the bubble diagrams as real scattering processes \cite{CN65}. In particular the internal lines represent real particles moving forward in time. Therefore the preparation-measurement pair in the $S$-matrix interpretation of quantum theory is related to the discontinuity $S_BS^{-1}_QS_A$, where $A$ is the preparation, $B$ is the measurement, and $Q$ is the observed quantum system.
\section{Bell Inequality}\label{BELL}
Consider the EPRB experimental setup where Alice has the choice of measurements $M_A=(a,a')$ and Bob has $M_B=(b,b')$. Each measurement has outcomes $+1$ or $-1$. Suppose that the entanglement shared by Alice and Bob are generated by virtual scattering as in the main text, which can be measured\footnote{We do not go into the mathematical details of the measure-theoretic checks. One can think of the path integral representation of the scattering amplitudes as a mental picture.} by a non-vanishing scattering amplitude $M$, e.g.
\[\vcenter{
\Qcircuit @C=1em @R=.7em {
& \multigate{1}{S} & \qw &{i,i'} \\
& \ghost{S}& \qw &{j,j'}
}}\quad\quad\Longrightarrow\quad M(ii'jj')\]
where $(ii'jj')$ are possible outcomes of measurements $(a_i,a'_{i'},b_j,b'_{j'})$. If each pair $(ij)$ has a joint probability  regardless of their compatibility, then  it is simply the marginal probability
\[
\sum_{i'j'}\quad\vcenter{
\Qcircuit @C=1em @R=.7em {
& \multigate{1}{S} & \qw &{i,i'} \\
& \ghost{S}& \qw &{j,j'}
}}\quad=\quad
\vcenter{
\Qcircuit @C=1em @R=.7em {
& \multigate{1}{S} & \qw &{i} \\
& \ghost{S}& \qw &{j}
}}\quad\quad\Longrightarrow\quad M(ij).
\]
Here $M(ij)$ is a new measure measuring the quantum entanglement between $(ij)$.

Now the Bell inequality can be readily obtained algebraically (see, e.g. \cite{Bell}). Define the correlation functions in analogy to those in path integrals as weighted sums, e.g.
\begin{equation}\label{mmm}
X(a,b)=\sum_{ij}i\cdot j\cdot M(ij).
\end{equation}
Then since $ij+i'j+ij'-i'j'=(i+i')j+(i-i')j'\leqslant2$, one has
\[X(a,b)+X(a',b)+X(a,b')-X(a',b')=\sum_{ii'jj'}(ij+i'j+ij'-i'j')M(ii'jj')\leqslant2.\]
The absolute value sign can be added to get the Bell inequality.

Based one the new measure of entanglement $M(ij)$, we can envision a second level of virtual scattering: the scattering amplitudes are those correlation functions as in \eqref{mmm}. So diagrammatically, we can represent them by bubble diagrams with the new interpretation that to each line is assigned an outcome such that $+1$ means the rightward-directness and $-1$ the leftward-directness. Now a plus sign in the bubble means that the product e.g. $i\cdot j=+1$ and a minus sign means $-1$. Suppose furthermore that when recombining the cluster-decomposed $S$-matrices, the signs in the bubbles are added together. Then one has
\[\xymatrix{
\sum_{(ij),(i'j),(ij'),(-i'j')}\ar@{-}[r] &*+[o][F-]{\pm}&\ar@{-}[l]&=&\ar@{-}[r] &*+[F]{ij+i'j+ij'-i'j'}&\ar@{-}[l]}\]
which reduces to the algebraic derivation.

\section{Preservation of von Neumann Entropy}\label{BBB}
Let $\rho_{\mathsf{s}}$ be a prepared quantum state, and $\Phi$ a CP map that describes the evolution of $\rho_{\mathsf{s}}$ in the classical environment. Suppose that $\Phi$ is trace-preserving and unit-preserving, namely, a bistochastic quantum operation. 

Recall  that a bistochastic quantum operation preserves the von Neumann entropy $S(\Phi(\rho))=S(\rho)$ iff it is unitary $\Phi^\dagger\Phi={\bf1}$. In \cite{ZW11} it is shown that this condition is equivalent to the following:
the quantum state $\rho$ can be written as
\[\rho=\bigoplus_k\rho_k=\bigoplus_kp_k\rho_k^L\otimes\frac{1}{d_k^R}{\bf 1}^R\]
where $\sum_kp_k=1$, $d_k=\dim\mathcal{H}_k$; the bistochastic quantum operation $\Phi$ can be decomposed as
\[\Phi=\bigoplus_k\Phi_k=\bigoplus_k\text{Ad}_{U_k}\otimes\Phi_k^R\]
where the $U_k$ are unitary operators. Now just by staring at the above decomposition formula of $\rho$ and $\Phi$, we find that the direct sum $\oplus$ corresponds to the sum in the cluster decomposition of an $S$-matrix,  the operations  $\text{Ad}_{U_k}$ on the $L$-states to the part of a bubble diagrams including nontrivial bubbles (i.e. with at least two incoming/ougoing lines), and likewise $R$-states as identities to the lines with trivial bubbles. Moreover, the probabilities $p_k$ are weights of topologically different digrams with specified lines and bubbles.
 
As an example, let us draw the bubble diagram for the cluster decomposition of the $S$-matrix for a $3\rightarrow3$ scattering 
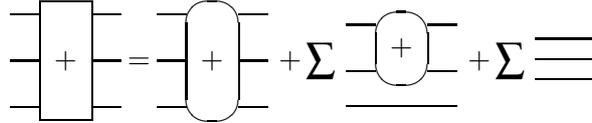
\[\vcenter{\Qcircuit @C=1em @R=.7em {
& \multigate{2}{+} & \qw \\
& \ghost{+}& \qw \\
& \ghost{+} & \qw
}}=
\vcenter{\Qcircuit @C=1em @R=.7em {
& \multimeasure{2}{+} & \qw \\
& \ghost{+}& \qw \\
& \ghost{+} & \qw
}}~+\sum~
\vcenter{\Qcircuit @C=1em @R=.7em {
& \multimeasure{1}{+} & \qw \\
& \ghost{+}& \qw \\
& \qw & \qw
}}~+\sum~
\vcenter{\Qcircuit @C=1em @R=.7em {
& \qw & \qw \\
& \qw& \qw \\
& \qw & \qw
}}\]
The forms of the states can be read from the diagram in an obvious way
\begin{align*}
&\rho_1=p_1\rho^{3d}\otimes(...),\quad\rho_2=p_2\rho^{2d}\otimes{\bf1}^{1d},\quad\rho_3=p_3(...)\otimes\frac{{\bf 1}^{3d}}{3},\quad p_1+p_2+p_3=1
\end{align*}
where $(...)$ represents discarded part. Similarly, the quantum operations are 
\[\Phi_1=\text{Ad}_{U}^{3d}\otimes(...),\quad\Phi_2=\text{Ad}_U^{2d}\otimes\Phi^{1d},\quad\Phi_3=(...)\otimes\Phi^{3d}.\]

\bibliographystyle{amsalpha}

\textsc{Wuhan Institute of Physics and Mathematics, Chinese Academy of Sciences, Wuhan 430071, China;\\
University of  Chinese Academy of Sciences, Beijing 100049, China.}\\
{\sf E-mail}: kankuohsiao@whu.edu.cn; xkguo@wipm.ac.cn

\end{document}